\documentclass[10pt,conference]{IEEEtran}

\usepackage{mathrsfs}
\usepackage{psfrag}
\usepackage{graphicx}
\usepackage{color}
\usepackage{fancyhdr,epsfig}
\usepackage{amsthm}
\usepackage{dsfont,enumerate}

\def\wh{\widehat}

\def\mb{\mathbb}
\def\m{\mathcal}
\def\md{\mathds}
\def\eps{\varepsilon}
\def\tn{\textnormal}

\newcommand{\dfn}{ \stackrel{\tn{def}}{=} }

\newcommand {\bx} {\mbox{\boldmath $x$}}
\newcommand {\by} {\mbox{\boldmath $y$}}

 \usepackage{amsmath, amssymb}

\newtheorem{lemma}{Lemma}

\newtheorem{corollary}{Corollary}
\newtheorem{theorem}{Theorem}

\newtheorem{remark}{Remark}

\allowdisplaybreaks[3]

\begin{document}

\IEEEoverridecommandlockouts

\title{Searching with Measurement Dependent Noise}

\author{Yonatan~Kaspi, Ofer~Shayevitz and Tara~Javidi\thanks{Y. Kaspi and T. Javidi are with the Information Theory and Application (ITA) Center at the University of California, San Diego, USA. O. Shayevitz is with the Department of EE--Systems, Tel Aviv University, Tel Aviv, Israel. Emails: \{ofersha@eng.tau.ac.il, ykaspi@ucsd.edu, tjavidi@ucsd.edu\}. The work of O. Shayevitz was partially supported by the Marie Curie Career Integration Grant (CIG), grant agreement no. 631983.}}

\maketitle

\begin{abstract}
Consider a target moving with a constant velocity on a unit-circumference circle, starting from an arbitrary location. To acquire the target, any region of the circle can be probed for its presence, but the associated measurement noise increases with the size of the probed region. We are interested in the expected time required to find the target to within some given resolution and error probability. For a known velocity, we characterize the optimal tradeoff between time and resolution (i.e., maximal rate), and show that in contrast to the case of constant measurement noise, measurement dependent noise incurs a multiplicative gap between adaptive search and non-adaptive search. Moreover, our adaptive scheme attains the optimal rate-reliability tradeoff. We further show that for optimal non-adaptive search, accounting for an unknown velocity incurs a factor of two in rate. 
\end{abstract}

\section{Introduction}
Suppose a point target is arbitrarily placed on the unit-circumference circle. The target then proceeds to move at some constant velocity $v$ (either known or unknown). An agent is interested to determine the target's position and velocity to within some resolution $\delta$, with an error probability at most $\eps$, as quickly as possible. To that end, the agent can probe any region of his choosing (contiguous or non-contiguous) on the circle for the presence of the target, say once per second. He then receives a binary measurement pertaining to the presence of the target in the probed region, which is corrupted by additive binary noise. While the noise sequence is assumed to be independent over time, its magnitude will generally depend on the size of the probed region. This postulate is practically motivated if one imagines that the circle is densely covered by many small sensors; probing a region then corresponds to activating the relevant sensors and obtaining a measurement that is a (Boolean) function of the sum of the noisy signals from these sensors. We therefore further operate under the assumption that the larger the probed region, the higher the noise level. Our goal is to characterize the relation between $\eps$, $\delta$, and the expected time $\mb{E}(\tau)$ until the agent's goal is met, for both adaptive and non-adaptive search strategies. 

The case of stationary target search with measurement independent noise $p$ is well known (see e.g. \cite{burnashev1974interval}) to be equivalent to the problem of channel coding with noiseless feedback over a Binary Symmetric Channel (BSC) with crossover probability $p$, where the message corresponds to the target, the number of messages pertains to inverse of the resolution, the channel noise plays the role of measurement noise, and the existence of noiseless feedback pertains to the fact that the agent may use past measurements to adapt his probing strategy. Based on the results of \cite{burnashev_exp} it can be readily shown that using adaptive strategies one can achieve 
\begin{equation*}
\mb{E}(\tau) = \frac{\log{(1\slash\delta)}}{C(p)} + \frac{\log{(1\slash\eps)}}{C_1(p)} + \mathrm{O}(\log\log{\frac{1}{\delta\eps}})
\end{equation*}
where $C(p)$ is the Shannon capacity of the BSC with crossover probability $p$, and $C_1(p) = D(p\|1-p)$. This result is also the best possible up to sub-logarithmic terms. For non-adaptive strategies, standard channel coding results \cite{GallagerBook} indicate for any fixed $0<R<C(p)$ there exists a strategy such that 
\begin{equation*}
\tau = \frac{\log{(1\slash\delta)}}{R},\qquad \log{(1\slash\eps)} = \frac{E(R,p)}{R}\cdot \log{(1\slash\delta)}
\end{equation*}
where $E(R,p)$ is the reliability function of the BSC, for which bounds are known \cite{GallagerBook}. Hence, the minimal expected search time (with a vanishing error guarantee) is roughly the same for adaptive and non-adaptive strategies in the limit of high resolution $\delta\to 0$, and is given by $\mb{E}(\tau) \approx \frac{\log{(1\slash\delta)}}{C(p)}$. This directly corresponds to the fact that feedback does not increase the capacity of a memoryless channel \cite{shannon_zero_error}. Adaptive search strategies do however exhibit superior performance over non-adaptive strategies for a fixed resolution, attaining the same error probability with a lower expected search time. They are also asymptotically better if a certain exponential decay of the error probability is desired, which directly corresponds to the fact that the Burnashev exponent \cite{burnashev_exp} exceeds the sphere packing bound \cite{GallagerBook} at all rates below capacity. 

The contribution of this work is threefold: 
\begin{itemize}[leftmargin=*]
\item In contrast to the case of measurement independent noise, it is shown that for known velocity and measurement dependent noise there exists a multiplicative gap between the minimal expected search time for adaptive vs. non-adaptive strategies, in the limit of high resolution. This \textit{targeting rate} gap generally depends on the variability of the measurement noise with the size of the probed region, and can be arbitrarily large. The source of the difference lies mainly in the fact that from a channel coding perspective, the channel associated with measurement dependent noise is time-varying in quite an unusual way; it depends on the choice of the \textit{entire codebook}. The maximal targeting rates achievable using adaptive and non-adaptive strategies under known velocity are given.  
\item A rate-reliability tradeoff analysis is provided for the proposed adaptive and non-adaptive schemes, under known velocity. It is shown that the former attains the best possible tradeoff. 
\item For unknown velocity, the maximal targeting rate achievable using non-adaptive schemes is shown to be reduced by a factor of two relative to the case of known velocity.  
\end{itemize}


\section{Preliminaries}
\subsection{Notations}
The Shannon entropy of a random variable (r.v.) $X$ is denoted by $H(X)$. The mutual information between two jointly distributed r.v.s $X$ and $Y$ is denoted $I(X;Y)$. A BSC($p$) is a BSC with crossover probability $p$.  When $X\sim\textrm{Bern}(q)$ and $Y$ is the output of a BSC($p$) with input $X$, we write $I(q,p)$ for $I(X;Y)$. We write $C(p)$ for the Shannon capacity of a BSC($p$), and $C_1(p)$ for the relative entropy $D(p\|1-p)$. The cardinality of a finite set $S$ is denoted by $|S|$. The Lebesgue measure of a set $S\subset\mb{R}$ is similarly denoted by $|S|$. We write $\mathds{1}(\cdot)$ for the indicator function. The \textit{cyclic distance} between $a,b\in[0,1)$ is the associated angular distance on the unit-circumference circle, i.e., $|a-b|_c \dfn \min\{|a-b|,1-|a-b|\}$.  
\subsection{Setup}
Let $w_0\in[0,1)$ be the initial position of the target, arbitrarily placed on the unit interval\footnote{For simplicity of notation, we will think of the target as moving on the unit interval modulo 1, instead of on the circle.}. The target moves at a fixed but unknown velocity $v\in[0,1)$, i.e., at time $n$ the position of the target is given by
\begin{equation*}
w_n = w_0 + v\cdot n \quad (\textrm{mod }1)
\end{equation*}     

At time $n$, the agent may seek the target by choosing (possibly at random) any measurable \textit{query set} $S_n\subset [0,1)$ to probe. Without loss of generality, we will assume throughout that $|S_n| \leq \frac{1}{2}$ almost surely. Let $X_n = \md{1}(w_n\in S_n)$ denote the clean binary signal indicating whether the target is in the probed region. The agent obtains a corrupted version $Y_n$ of $X_n$, with noise level that corresponds to the size of the region $S_n$. Specifically, 
\begin{align*}
Y_n = X_n + Z_n \;(\textrm{mod}\;2),  
\end{align*}
where $Z_n \sim \textrm{Bern}\left(p[|S_n|]\right)$, and where $p:(0,1/2]\mapsto [0,1/2)$ is a continuous and monotonically non-decreasing function.  

A \textit{search strategy} is a causal protocol for determining the sets $S_n = S_n(Y^{n-1})$, associated with a stopping time $\tau$ and estimators $\wh{W}_\tau = \wh{W}_\tau(Y^\tau),\wh{V}=\wh{V}(Y^\tau)$ for the last position and the velocity. A strategy is said to be \textit{non-adaptive} if the choice of the region $S_n$ is independent of $Y^{n-1}$, i.e., the sets we probe do not depend on the observations. In such a case, the stopping time is also fixed in advance. Otherwise, the strategy is said to be \textit{adaptive}, and may have a variable stopping time. A strategy is said to have \textit{search resolution} $\delta$ and error probability $\eps$ if for any $w_0,v$,
\begin{equation*}
\Pr(\max\{|\wh{W}_\tau-w_\tau|_c,|\wh{V}-v|_c\} \leq \delta) \geq 1-\eps
\end{equation*}
We are interested in the expected search time $\mb{E}(\tau)$ for such strategies, and specifically in the \textit{maximal targeting rate}, which is the maximal ratio $\frac{\log{1\slash\delta}}{\mb{E}(\tau)}$ such that $\eps\to 0$ is possible as $\delta\to 0$. We say that a sequence of strategies indexed by $k$ achieves a \textit{targeting rate} $R$ and an associated \textit{targeting reliability} $E = E(R)$, if $\delta_k\to 0$ as $k\to \infty$ and
\begin{equation*}
\mb{E}(\tau_k) \leq \frac{\log{(1\slash\delta_k)}}{R},\qquad \log{(1\slash\eps_k)} \geq \frac{E}{R}\cdot \log{(1\slash\delta_k)}
\end{equation*}
for all $k$ large enough. 


\section{Non-adaptive strategies}
We state our main result for the non-adaptive case. Both known and unknown velocities are treated. In our proofs we will assume the latter; the former simpler case follows easily. 
\begin{theorem}\label{thrm:non-adapt}
Let $p[\cdot]$ be a measurement noise function. For non-adaptive search strategies, the maximal targeting rate is given by 
  \begin{equation}\label{eq:rate_non_adaptive}
    \max_{q\in(0,\frac{1}{2})} \kappa I(q,p[q])
  \end{equation}
where $\kappa=\frac{1}{2}$ for an unknown velocity, and $\kappa=1$ for a known velocity. Moreover, for any $R$ below the maximal targeting rate, there exist non-adaptive search strategy such that 
\begin{equation*}
\tau = \frac{\log{((1\slash\delta))}}{R},\qquad \log{(1\slash\eps)} = \frac{E_r(R,q^*)}{R}\cdot \log{(1/\delta)}
\end{equation*}
where $q^*$ is the maximizer in \eqref{eq:rate_non_adaptive}, and 
\begin{equation*}
  E_r(R,q^*) = \max_{\rho\in (0,1)} E_0(\rho,q^*) - \rho R/\kappa
\end{equation*}
is the random coding exponent \cite{GallagerBook} for a  BSC$(p[q^*])$ with input distribution $q^*$, at rate $R/\kappa$. 
\end{theorem}
\subsection{Proof of Converse}
Denote the fixed stopping time by $\tau = N$. Let  $\{S_n\}_{n=1}^N$ be any non-adaptive strategy achieving an error probability $\eps$ with search resolution $\delta$. We prove the converse holds even under the less stringent requirement where the initial position and velocity are uniformly distributed $(W_0,V)\sim\textrm{Unif}([0,1)^2)$. Partition the unit interval into $\lceil\beta/\delta\rceil$ equi-sized intervals for some constant $\beta\in(0,\tfrac{1}{2})$, and let $W_N'$ be the index of the interval containing $W_N$. Similarly, let $V'$ be the index of the interval containing $V$. It is easy to see that the scheme $\{S_n\}$ can be made to return $W_N'$ with error probability at most $\eps'\dfn \eps + 4\beta(1-\beta)$, where the latter addend stems from the probability that $(\wh{W}_N,\wh{V})$ is too close to a boundary point. 

Note that $X_n\sim\mathrm{Bern}(q_n)$ where $q_n\dfn |S_n|$ and that $Y_n$ is obtained from $X_n$ through a memoryless binary symmetric channel with a time-varying crossover probability $p\left[q_n\right]$. Following the steps of the converse to the channel coding theorem, we have
\allowdisplaybreaks{\begin{IEEEeqnarray}{rCl}\label{eq:converse}
2\log\left(\frac{\beta}{\delta}\right) &=& H(W_N',V') \notag\\ 
&=& I(W_N',V';Y^N) + H(W_N',V'|Y^N) \notag\\ 
&\stackrel{(\textrm{a})}{\leq}& I(W_N',V';Y^N) + N\eps' \notag\\ 
&=& \sum_{n=1}^N I(W_N',V';Y_n|Y^{n-1}) + N\eps' \notag\\ 
&\leq& \sum_{n=1}^N I(W_N',V',Y^{n-1};Y_n) + N\eps' \notag\\ 
&\leq& \sum_{n=1}^N I(W_N',V',W_n,Y^{n-1};Y_n) + N\eps' \notag\\ 
&\stackrel{(\textrm{b})}{=}& \sum_{n=1}^N I(X_n;Y_n) + N\eps' \notag\\ 
&\stackrel{(\textrm{c})}{=}& \sum_{n=1}^N I(q_n,p[q_n]) + N\eps', 
\end{IEEEeqnarray}}
where (a) is by virtue of Fano's inequality, (b) follows since $X_n$ is a function of $W_n$ and the measurement noise is independent across time, and (c) stems from the fact that the  crossover probability sequence $p[q_n]$ is a fixed (time-varying) function of the codebook. Note that here, in contrast to the standard memoryless channel coding setup where the channel noise is strategy independent, (b) above does not generally hold when an adaptive strategy (i.e., feedback) is employed; this stems from the fact that in this case, the intensity of the observation noise would generally depend on $Y^{n-1}$, and therefore $Y_n-X_n-Y^{n-1}$ would not form a Markov chain. Dividing by $N$ we obtain 
\begin{IEEEeqnarray*}{rCl}
R &=& \frac{\log(1/\delta)}{N} \leq  \frac{1}{2N}\left(\sum_{n=1}^N I(q_n,p[q_n]) - 2\log\beta\right) + \frac{\eps'}{2} \\
&\leq&  \frac{1}{2}\left(\sup_{q\in(0,\frac{1}{2})} I(q,p[q]) -\frac{2\log\beta}{N}+ \eps + 4\beta(1-\beta)\right).
\end{IEEEeqnarray*}
Noting that the inequality above holds for any $\beta\in(0,\tfrac 1 2)$, the converse now follows by taking the limit $N\to\infty$, and then requiring $\eps\to 0$. 

\subsection{Proof of Achievability}
Achievability is obtained via random coding using an input distribution $q^*$ that achieves the supremum in \eqref{eq:rate_non_adaptive}. A sketch of the proof is now given. We partition the unit interval into $M = \frac{N}{\delta}$ equi-sized subintervals $\{b_m\}$. Each pair of initial position and velocity $(w_0,v)$ naturally induces a \textit{trajectory} $m(w_0,v)$ w.r.t. this partition. We say that two trajectories $m(w_0,v)$ and $m(w_0',v')$ are \textit{$(\delta,N)$-close} if $|w_0-w_0'|_c\leq 
\delta$ and $|v-v'|_c\leq \frac{\delta}{N}$. Otherwise, we say the trajectories are $(\delta,N)$-far.   
\begin{lemma}\label{lem:trajec}
  The number of different trajectories is upper bounded by $K = M^2\cdot \mathrm{O}(\textrm{poly}(N))$. Moreover, if two trajectories intersect more than once, then their corresponding initial positions and velocities are $(\delta,N)$-close.  
\end{lemma}

We now draw a codebook with $M$ rows, where each row $\bx_m$ has $N$ bits. The codebook is drawn i.i.d. $\textrm{Bern}(q^*)$. We define our random query set $S_n$ according to the codebook's columns:  
\begin{align*}
  S_n \dfn A + Bn + \bigcup_{m:x_{m,n}=1}b_m 
\end{align*}
where $(A,B)\sim\textrm{Unif}([0,1)^2)$ serve as a random ``dither'', mutually independent of the measurement noise. This dithering procedure renders our setting equivalent to the setup where the initial position and velocity are uniform and independent, and where the query sets are given by $A=B=0$. We shall proceed under this latter setup. 

The next Lemma stems directly from Chernoff's bound.  
\begin{lemma}
Let $\m{A}$ be the event where $||S_n|-q^*|\leq \epsilon$ for all $n$. Then for any $\epsilon > 0$, $\Pr(\m{A}^c) = 2^{-2^{O(N)}}$. 
\end{lemma}
\begin{remark}
Under the event $\m{A}$, we can safely assume that the measurements are observed through a BSC$(p[q^*+\epsilon])$, since we can always artificially add noise to the observations at any time $n$ for which $|S_n| < q^*+\epsilon$.   
\end{remark}

Our codebook induces a set of \textit{trajectory codewords} $\{x_{m(w_0,v),n}\}_{n,w_0,v}$. Note that each trajectory codeword corresponds to a set of possible initial positions and velocities. With a slight abuse of notations, we denote the trajectory codewords by $\{\bx_k\}_{k=1}^K$. After $N$ queries, we find the trajectory codeword that has the highest likelihood under the assumption that the measurements are observed through a BSC$(p[q^*+\epsilon])$. We now show that the likelihood of the correct trajectory codeword is with high probability higher than that of all trajectory codewords whose associated initial position or velocity are at least $(\delta,N)$-far. Hence, the initial position and velocity of the decoded trajectory will be $(\delta,N)$-close to the correct one, with high probability.  Note that if the target had been stationary, we would have searched for the highest likelihood row just as in channel coding.

We write the average probability of error as 
\begin{equation*}
\overline{P}_e = \Pr(\m{A})\Pr(e|\m{A}) + \Pr(\m{A}^c)\Pr(e|\m{A}^c).  
\end{equation*}
The second term vanishes double exponentially fast. For the other term we have
\begin{align*}
 	\Pr(e|\m{A})=\sum_{\bx_k}\Pr(\bx_k|\m{A})P_{\m{A}}(\by|\bx_k)\Pr(e|\bx_k,\by,\m{A}),
\end{align*}
where $\by$ are the noisy observations and $P_{\m{A}}(\by|\bx_m)$ is the $BSC(q^*+\epsilon)$ induced by the event $\m{A}$ (and possible randomization). Let $\m{E}_{k'}$ denote the event that the trajectory codeword $\bx_{k'}$ is chosen instead of $\bx_k$. Let $T_k$ be the set of all $k'$ for which either the velocity or the initial position of each of the trajectories associated with $\bx_k'$, are more than  $\delta$-far from those of $\bx_k$. 
\begin{align}\label{eq:Tk}
 	\Pr(e|\bx_k,\by,\m{A}) \leq \sum_{k'\in T_k}\Pr(\m{E}_{k'}|\m{A})
\end{align}
and 
\begin{align}
 	\Pr(\m{E}_{k'}|\m{A}) = \sum_{\bx_{k'}: P_{\m{A}}(\by|\bx_k)\leq  P_{\m{A}}(\by|\bx_{k'})} \Pr(\bx_{k'}|\bx_k,\m{A})\label{eq:CondErr}
\end{align}
Note that unlike \cite[eq. 5.6.8]{GallagerBook}, we cannot assume the trajectory codewords are independent under event $\m{A}$. Furthermore, for $k'\in T_k$ the trajectories may intersect once. We therefore have that $\Pr(\bx_k, \bx_{k'}|\m{A}))\leq \frac {\Pr(\bx_k,\bx_{k'})}{1-\Pr(\m{A}^c)} \leq \frac {Q(\bx_k)Q(\bx_{k'})}{(1-\Pr(\m{A}^c))q_{min}}$ and $\Pr(\bx_k|\m{A})\geq Q(\bx_k) - \Pr(\m{A}^c)$, where $Q(\cdot)$ denotes the random coding prior, and $q_{min}$ denotes the probability of the least probable binary symbol under Q. Using this and Bayes rule, for $N$ large enough we have:
\begin{align}
&P(\bx_{k'}|\bx_k,\m{A}) \leq \frac{Q(\bx_k)Q(\bx_{k'})}{(1-\Pr(\m{A}^c))(Q(\bx_k)-\Pr(\m{A}^c))q_{min}} \notag
\\ &\leq \frac{Q(\bx_{m'})}{(1-\Pr(\m{A}^c)/q_{min}^N)^2q_{min}} = \frac{Q(\bx_{m'})}{q_{min}}(1+2^{-2^{O(N)}})\label{eq:CondProb}
\end{align}
After substituting \eqref{eq:CondProb} in \eqref{eq:CondErr} and \eqref{eq:Tk} and plugging in $\delta=2^{-NR}$, we can follow Gallager's derivation of the random error exponent \cite{GallagerBook} almost verbatim, with the following two distinctions: 1) By Lemma \ref{lem:trajec} the effective number of messages is now $|T_k| = K = M^2\cdot\mathrm{O}(\textrm{poly}(N))$; and 2) for any finite $N$, the exponent is multiplied by a constant pertaining to the double exponential penalty and to $q_{min}$, but this constant converges to unity as $N$ grows. The exponent is positive as long as $R\leq I(q^*,p[q^*+\epsilon])/2$. As $\epsilon$ is arbitrary, this concludes the proof of achievability.

\section{Adaptive strategies}
In this section, we consider the gain to be reaped by allowing the search decisions to be made adaptively. For simplicity, we assume here that the velocity is known in advance, and hence without loss of generality can be assumed to be zero. We will again use dithering to make the initial position appear uniformly random. Here, the duration of search $\tau$ will generally be a random stopping time dependent on the measurements sample path. Moreover, the choice of probing regions $S_n$, for $n$ up to the horizon $\tau$, 
can now depend on past measurements. We characterize this gain in terms of the maximal targeting rate, and the targeting rate-reliability tradeoff. As we shall see, adaptivity allows us to achieve the maximal possible rate and reliability, i.e., those associated with the minimal observation noise $p[0]$. 

\subsection{Non Adaptive Search with Validation}
As a first attempt at an adaptive strategy, we continue with the non-adaptive search from the previous section, but allow the agent to validate the outcome of the search phase. We will consider two validation schemes, due to Forney  \cite{Forney1968} and Yamamoto-Itoh \cite{YamaItoh1980}. 

\begin{figure}[htp]
\centering
\includegraphics[width=0.45\textwidth]{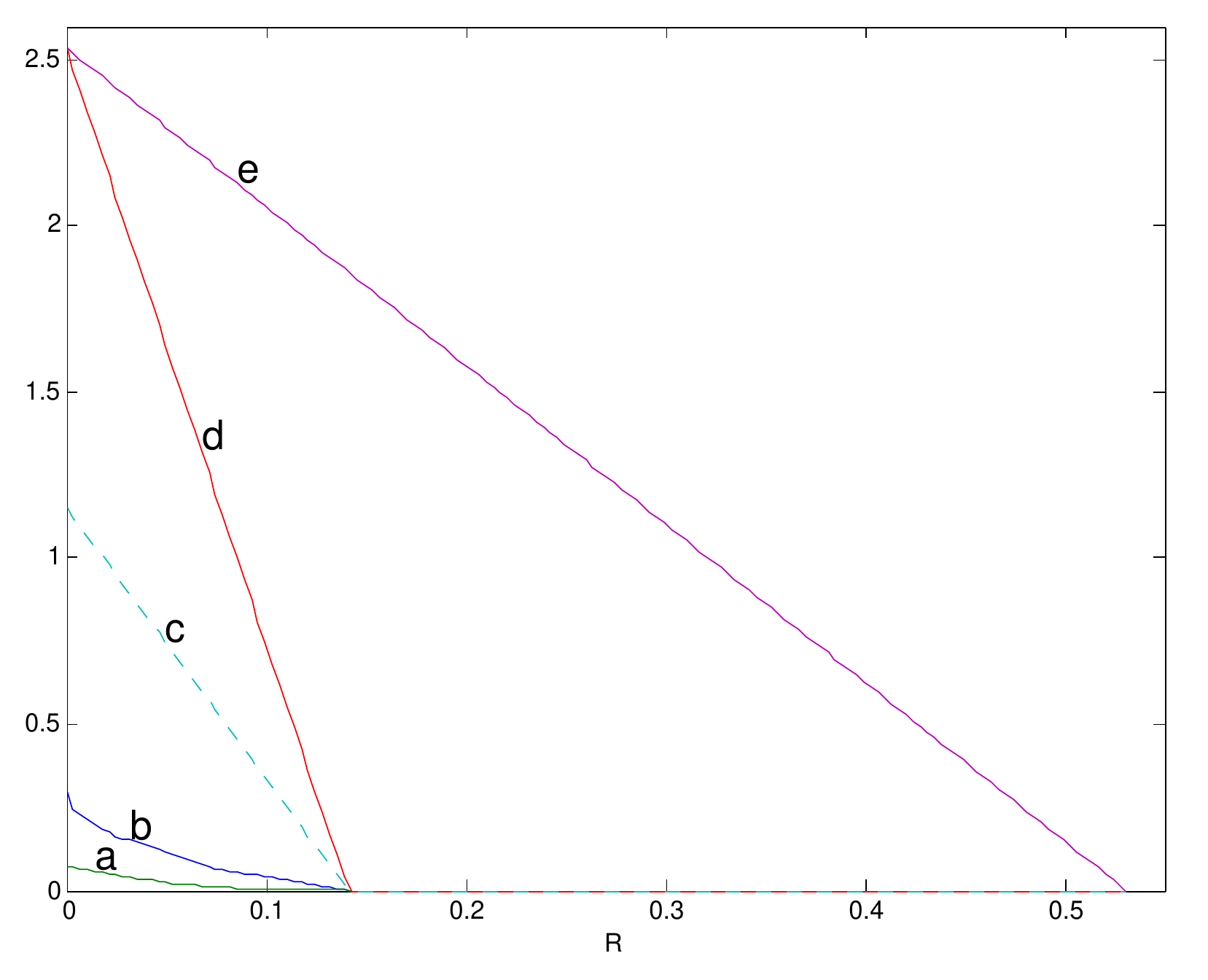} 
\caption{Error exponents (known velocity) for noise growing linearly with size: $p[0]=0.1, p[\frac 1 2]=0.45$ (a) Random coding (b) Decision feedback (c) Burnashev's upper bound for BSC$(p[q*])$ (d) Yamamoto-Itoh validation for the non-adaptive scheme (e) Yamamoto-Itoh validation for BSC$(p[0])$}.\label{Fig:Exponents}
\end{figure}

In \cite{Forney1968}, Forney considered a communication system in which a decoder, at the end of the transmission can signal the encoder to either repeat the message or continue to the next one. Namely, it is assumed that a one bit ``decision feedback'' can be sent back to the transmitter at the end of each message block. This is achieved by adding an erasure option to the decision regions, that allows the decoder/agent to request a ``retransmission'' if uncertainty is too high, i.e., to restart the exact same coding/search process from scratch. More concretely, given $Y^N$, a codeword $k$ will be declared as the output if $\frac {P(y^N|\bx_k)} {\sum_{k'\neq k}P(y^N|\bx_k')}\geq 2^{NT}$, where $T>0$ governs the tradeoff between the probability of error and the probability erasure. Let $\m{E}$ denote the event of erasure. The expected search duration will be $\frac N {1-\Pr(\m{E})}$. While having negligible effect on the rate (as long as $\Pr(\m{E})$ vanishes as N grows), the results of  \cite{Forney1968} immediately imply that such a scheme drastically improves the error exponent compared to non-adaptive schemes (see Fig.\ref{Fig:Exponents}). 

The second validation scheme we consider was proposed by Yamamoto and Itoh in \cite{YamaItoh1980} in the context of channel coding with clean feedback. Unlike Forney's scheme which requires only one bit of feedback, this scheme requires the decoder to feed back its decision. While perfect feedback is impractical in a communication system, in our model it is inherent and can be readily harnessed. After completing the search phase with resolution $\delta$, the agent continues to probe the estimated target location, namely an interval of size $\delta$. If the probed region contains the target, the output of the validation phase should look like a sequence of '1's passing through a BSC$(p[\delta])$. Thus, if the validation output is typical w.r.t. to a binary source with $\Pr('1')=1-p[\delta]$, the agent outputs that region as the final decision. Otherwise, the whole search is repeated from scratch. Specifically, After the $N$ queries of the non-adaptive search, we probe the aforementioned region $\lambda N$ more times, where $0\leq \lambda\leq \infty$ determines the tradeoff between rate and reliability.  Let $\m{E}$ denote the event that the search is repeated. This happens if the wrong region has been chosen, or otherwise if the observations in the validation step were not typical. Both these events will have vanishing probabilities and therefore the rate will be negligibly affected; the average search length is now $\mb{E}(\tau) = \frac {N(1+\lambda)}{1-\Pr(\m{E})}$. Following the derivations of \cite{YamaItoh1980} with $\lambda=\frac {I(q^*;p[q^*])} {R} -1$, and noting that $\delta$ can be made arbitrarily small, we obtain: 
\begin{lemma}
The targeting rate-reliability tradeoff for non-adaptive scheme with a Yamamoto-Itoh validation is given by
\begin{equation*}
E = C_1(p[0])\cdot \left(1-\frac{R}{I(q^*;p[q^*])}\right)  
\end{equation*}  
\end{lemma}
Note that with this search strategy, we get better reliability than the optimal one for the $BSC(q^*)$ with feedback (given by Burnashev \cite{burnashev1974interval}) since the validation is done over the least noisy channel (see Fig.\ref{Fig:Exponents}).

\subsection{Two-Phase Search with Validation}
In this section, we show that a simple two-phase scheme with validation achieves the best possible performance, improving upon non-adaptive strategies (with and without validation) both in maximal targeting rate and in targeting rate-reliability tradeoff. 
      
\begin{theorem}
Let $p[\cdot]$ be a measurement noise function. For any $\alpha\in (0,\tfrac{1}{2})$, there exists a search scheme with error probability $\eps$ and resolution $\delta$, satisfying   
\begin{equation*}
\label{binC}
\mathbb{E} [\tau]\le  \left(\frac{\log(1\slash\alpha)}{C(p[q^*])} + \frac{\log (1\slash\delta)}{C(p[\alpha])} + \frac{\log(1\slash\epsilon)}{C_1(p[\delta])}\right)\left(1+\mathrm{o}(1)\right).
\end{equation*}
\end{theorem}
\begin{corollary}
By letting $\alpha$ vanish much slower than $\delta$, we conclude that the maximal targeting rate for adaptive schemes is given by 
  \begin{equation*}
    C(p[0]) \dfn \max_{q\in(0,\frac{1}{2})} I(q,p[0])= I(\tfrac{1}{2}, p[0]),
  \end{equation*}
which is the capacity of the least noisy BSC associated with the measurements, which is the best possible. The associated targeting rate-reliability tradeoff is 
\begin{equation*}
  E(R) =  C_1(p[0])\left(1 -\frac{R}{C(p[0])}\right).
\end{equation*}
which is also the best possible.   
\end{corollary}
\begin{remark}
  Juxtaposing Theorem \ref{thrm:non-adapt} and the Corollary above, we conclude that (unlike the case of constant interval-independent noise) adaptive search strategies outperform the optimal non-adaptive strategy in both targeting rate and reliability.  
\end{remark}
\begin{proof}
We prove the theorem for a fixed $\alpha$ and $\delta,\eps\to 0$. In the first search phase, the agent employs the optimal non-adaptive search strategy with $\tau=\log{N}$ and resolution $\alpha$, i.e. with a vanishing rate $R = \frac{\log{1\slash\alpha}}{\log{N}}$. At the end of this phase, the agent knows an interval of size $\alpha$ containing the target with probability $1- \textrm{o}(N)$.

In the second phase, the agent ``zooms-in'' and performs the search only within the interval obtained in the first phase. To that end, the agent employs the optimal non-adaptive search strategy with $\tau=\lambda N-\log{N}$ and resolution $\delta = 2^{-(\lambda N-\log{N})R}$, i.e. with  rate $R = \frac{\log{1\slash\delta}}{\lambda N-\log{N}}$, with the query sets properly shrunk by a factor of $\alpha$. We note that in this phase, all queried sets are of size smaller than $\alpha/2$, hence the associated noise is less that $p[\alpha]$. Therefore, if the rate $R< C[p[\alpha]]$, then at the end of this phase the agent knows an interval of size $\delta$ containing the target with probability $1- \textrm{o}(N)$. 

At this point, the agent perform the Yamamoto-Itoh validation step of length $(1-\lambda)N$, which queries a fixed interval of size $\delta$. If not successful, the agent repeats the whole two-phase search from scratch. The expected stopping time of this procedure is $\frac{N}{1-o(N)}$, and the error probability  decays exponentially with an exponent controlled by trading off the search and validation as before, yielding the associated Burnashev behavior for the channel $p[\delta]$. 
\end{proof}

\section{Conclusions and Further Research}
In this paper, we considered the problem of acquiring a target moving with known/unknown velocity on a circle starting from an unknown position, under the physically motivated observation model where the noise intensity increases with the size of the queried region. For a known velocity, we showed that unlike the constant noise model, there can be a large gap in performance (both in targeting rate and reliability) between adaptive and non-adaptive search strategies. The various rate-reliability tradeoffs discussed herein are depicted in Fig. \ref{Fig:Exponents}.  Furthermore, we demonstrated that the cost of accommodating an unknown velocity in the non-adaptive setting, is a factor of two in the targeting rate, as intuition may suggest.  

One may also consider other search performance criteria, e.g., where the agent is cumulatively penalized by the size of either the queried region or its complement, according to the one containing the target. The rate-optimal scheme presented herein, which is based on a two-phase random search, may be far from optimal in this setup. In such cases we expect that sequential search strategies, e..g, ones based on posterior matching \cite{Shayevitz11,naghshvar2013extrinsic}, would exhibit superior performance as they naturally shrink the queried region with time.

Other research directions include more complex stochastic motion models, as well as searching for multiple targets (a ``multi-user'' setting). For the latter, preliminary results indicate that the gain reaped by using adaptive strategies vs. non-adaptive ones diminishes as the number of targets increases.   
 
\section{Acknowledgement}
The authors would like to thank an anonymous reviewer for throughly reading the paper and for many useful comments. 
\bibliographystyle{IEEEtran}
\bibliography{Bib-1}

\end{document}